\documentclass[11pt,reqno]{amsart}
\usepackage{amsmath,amsfonts,amscd,amssymb}
\usepackage{amsmath, amsthm, amsopn, amssymb, graphicx, enumerate, geometry, afterpage, caption, subcaption, color, textcomp, bbm,bm}
\usepackage[bbgreekl]{mathbbol}

\usepackage{amsmath,amsfonts,amsthm,amsbsy,amssymb,dsfont,stmaryrd}
\usepackage{subcaption, bbm}
\usepackage[dvipsnames]{xcolor}
\usepackage{braket}

\date{January 16, 2023}
\usepackage{mathtools}

\numberwithin{equation}{section}

\colorlet{mylinkcolor}{violet}
\colorlet{mycitecolor}{YellowOrange}
\colorlet{myurlcolor}{Aquamarine}

\usepackage[outline]{contour}
\contourlength{1.2pt}
\usepackage{graphicx}
\usepackage{caption}
\makeatletter
\def\parsept#1#2#3{%
	\def\nospace##1{\zap@space##1 \@empty}%
	\def\rawparsept(##1,##2){%
		\edef#1{\nospace{##1}}%
		\edef#2{\nospace{##2}}%
	}%
	\expandafter\rawparsept#3%
}
\makeatother

\theoremstyle{plain}
\newtheorem{thm}{\protect\theoremname}[section]
\theoremstyle{plain}
\newtheorem{lem}[thm]{\protect\lemmaname}
\theoremstyle{plain}
\newtheorem{cor}[thm]{\protect\corollaryname}
\theoremstyle{plain}
\newtheorem{prop}[thm]{\protect\propositionname}
\newtheorem*{theorem-non}{Proposition}

\theoremstyle{remark}

\theoremstyle{plain}

\theoremstyle{remark}
\theoremstyle{plain}

\providecommand{\assumptionname}{Assumption}
\providecommand{\claimname}{Claim}
\providecommand{\corollaryname}{Corollary}
\providecommand{\definitionname}{Definition}
\providecommand{\lemmaname}{Lemma}
\providecommand{\propositionname}{Proposition}
\providecommand{\remarkname}{Remark}
\providecommand{\theoremname}{Theorem}
\providecommand{\examplename}{Example}
\providecommand{\conjecturename}{Conjecture}
\providecommand{\notationname}{Notation}

\def\be{\begin{equation}}
\def\ee{\end{equation}}
 
\def\ba{\begin{align}}
	\def\ea{\end{align}}

\def\K{{\mathcal B}}    

\newcommand{\R}{\mathbb{R}}

\def\bra#1{\langle #1 \vert}

\def\ket#1{\vert#1\rangle}

\newtheorem*{lem*}{\protect\lemmaname}

\newcommand{\1}{\mathbbm{1}}

\newcommand{\dif}{\mathrm{d}}

\usepackage{environ}

\usepackage{hyperref}

\NewEnviron{malign}{%
	\begin{align}\begin{split}
			\BODY
	\end{split}\end{align}
}

\title[Ruminations on Matrix Convexity and Quantum SSA 
]  
{\Large Ruminations on Matrix Convexity and the\\ 
 Strong Subadditivity of Quantum Entropy (corr)
}         

\dedicatory{\hspace{5.5cm} \it  
To E.H. Lieb and M.B. Ruskai, in celebration   \\ 
\hspace{6cm}    of  
their 1973 proof of the SSA  }

\author{\large Michael Aizenman}
\address{M.A. : Departments of Physics and  Mathematics, Princeton University, Princeton, NJ 08544, USA, \qquad Weston Visiting Professor at the Weizmann Institute of Science, Israel}
\email{aizenman@princeton.edu}

\author{Giorgio Cipolloni}
\address{G.C. Princeton Center for Theoretical Science, Princeton University, Princeton, NJ 08544, USA}
\email{gc4233@princeton.edu}

\begin{document}

\begin{abstract}   

The familiar second derivative test for convexity,  combined with resolvent calculus, is shown to yield a useful tool for the study of convex matrix-valued functions.   We demonstrate the applicability of this approach on a number of  theorems in this field.  These include convexity principles which  play an essential role in the Lieb-Ruskai proof of the strong subadditivity of quantum entropy. \\  

{\color{red}   
\noindent Note: The published version of this paper includes a local mistake in Eq. (5.7).  This is amended here in line with the follow up \emph{Erratum and Addendum} (LMP 2024).  
%
 }
\end{abstract}

\def\uppercasenonmath#1{} 
\let\MakeUppercase\relax 
\maketitle

\section{Introduction} 
 
\subsection{A bit of background} \hfill \\[-2ex] 

A real valued function $f: (a,b) \to \R$ is said to be   convex on $n$-matrices  if and only if for any pair of self adjoint $n\times n$ matrices $A_0, A_1$ with spectrum in $(a,b)$, and for any $\lambda\in (0,1)$  the following holds 
\be \label{eq:conv_def} 
f((1-\lambda) A_0 + \lambda A_1) \le   (1-\lambda)  f(A_0)  + \lambda f(A_1) 
\ee  
in the sense of quadratic forms.   
We denote by $\K_n(a,b)$ the (convex) set of self adjoint matrices with spectrum in $(a,b)$ and, adapting the notation of \cite{Sim19}, by $\mathcal C_n(a,b)$  the class of the corresponding $n$-matrix convex functions, with $\mathcal{C}_\infty(a,b):=\cap_{n=1}^\infty \mathcal C_n(a,b) $.  The function $f$ is called concave if $-f$ is convex. 

Some of the familiar implications of convexity of a real-valued functions  extend directly to its matrix version.   Among those is the Jensen inequality that states that for any function $f$ which is convex over a convex domain $D$, and any  probability measure $\rho(dM)$ of a compact  support within $D$ 
\be \label{Jensen}
\int f(M) \rho(dM)  \geq  f\left(  \int M \, \rho(dM) \right) \,.
\ee  

However, the matrix version of convexity turns to be a far more confining notion.  In particular, even for $n=2$ only  quadratic functions meet the defining condition  with no restriction on the self adjoint matrices' spectrum~\cite{HanTom09}.   A similar phenomenon holds for monotonicity, as was discovered earlier by Charles Loewner for matrix monotone functions.   Among Loewner's remarkable contributions is the realization that upon limiting the requirement of matrix-monotonicity to matrices of spectrum confined to a subinterval of $\R$ one gets a richer class of functions.  He showed that any $f$ that is monotone over matrices with spectrum in $(a,b)$ is analytic over $(a,b)$, has an analytic extension off the real line with the Herglotz property, and  admits a corresponding integral representation~\cite{Loe34} (cf.~\cite{Sim19}). 

Building on Loewner's theory, F. Kraus~\cite{Kra36} and later J. Bendat and S. Sherman~\cite{BenShe55},  have developed a related characterization of matrix-convex functions.  Its explicit statement is found below  in Theorem ~\ref{thm_2} which presents the easy half of the statement.     
Fundamental role in their  studies was played by the monotonicity, and  implied properties, of Loewner's \emph{divided difference} matrices  $L = (L_{i,j})_{i,j=1}^k$, of matrix elements  
\be  
 L_{i,j}  =  \frac{f(x_i) - f(x_j)}{x_i- x_j}
 \quad \mbox{(interpreted as $f'(x_i)$ in case  $i=j$)} \, , 
\ee 
associated with collections of sites 
  $x_1< x_2 < ... < x_k \subset (a,b)$.  
The link between the two notions established in \cite{BenShe55, Kra36}  is that  $f$ is matrix convex  
if and only if for all (equivalently for some \cite{Uch10}) $y\in (a,b)$ the function $g(x) := (f(x) -f(y)) / (x-y) $ is matrix monotone over $(a,b)$.\\ 

Aside from the analytical challenges described above,  convexity  has  been of interest as a key tool for a myriad of variational problems.  
It is then natural that matrix convexity drew attention also in the theory of entropy functions of quantum statistical mechanics and   quantum information theory, and somewhat independently in  discussions of matrix networks.
 For such applications of convexity, the more relevant challenge has been to establish convexity statement for specific functions of interest.  
 
Examples of results whose development was driven by such applications are mentioned below.   
 Among those 
  is a theorem of Lieb~\cite{Lie73}, which proved  the Wigner-Yanase-Dyson conjecture~\cite{WigYan63} and played a key role in the   Lieb-Ruskai  proof of strong sub-additivity of entropy~\cite{LieRus73PRL, LieRus73}, a property commonly referred to as SSA.  
From another direction, matrix convexity showed up in the work of Anderson and Duffin~\cite{AndDuf69} on parallel sums (equivalently harmonic means), an extension of which can be found in  
the  Kubo-Ando \cite{KubAnd80} theory of operator means\footnote{  
Some of these developments have proceeded on parallel tracks: the  paper of Anderson and Duffin~\cite{AndDuf69}, where matrix concavity of parallel sums is noted and proven, has neither cited the earlier theory nor was its value recognized in early works on quantum entropy.   A contributing factor may have been that convexity was mentioned in ~\cite{AndDuf69} only in passing, in one (\#24) of many theorems, and was not mentioned in the work's summary.} 
  
Given how rich is the theory related to the above we refer the reader for further details to the 
recent and thorough book by B. Simon \cite{Sim19}.

\subsection{Outline of the paper's contents}\hfill \\[-2ex] 

Our ruminations on matrix convexity start with the discussion of a local criterion for concavity, and show how it allows short proofs of some of the known theorems, which were originally derived by other means.   
These include concavity of parallel sums, and from that a statement known as Lieb concavity, that plays a key role in different proof of the above mentioned properties of quantum entropy functionals.  
That discussion does not add to the proof plan which was laid out in the original work of E.H. Lieb and M.B. Ruskai~\cite{LieRus73}, beyond what may perhaps be received as a simplification in the derivation of the enabling results.  

The proof of SSA presented here passes through the Lieb-Ruskai concavity of the conditional entropy of a composite system as function of the state.  More explicitly,  in the notation that is explained in Section~\ref{sec:entropy},   $ S(1\vert 2)\equiv S(\rho_{12})-S(\rho_2)$ is a concave function of the state operator $\rho$ -- a statement which is also of independent interest~\cite{Han16}.

\section{Local test of convexity}  
\label{sec_gloloc}

While some convexity relations are perhaps better grasped through convexity's non local expressions, such  as the Jensen inequality \eqref{Jensen} and its non-abelian extensions (cf. \cite{Eff09, HanPed03}),    
a well known sufficient condition for it is the positivity of the second derivative (employed in different ways in~\cite{BroVas00, HanTom09, Kra36, Lie73}). 
Here we shall employ the following simply stated version of such a  criterion for matrix valued functions.  

\begin{prop}
\label{martingale_criterion}  
A  sufficient condition for $f:(a,b)\to \R$ to be in $\mathcal{C}_n(a,b)$ is that for any matrix   $M\in \K_n(a,b)$ and any bounded self adjoint $Q$ of equal rank  the matrix valued function   $f(M+tQ)$ is twice differentiable at $t=0$ and satisfies
\be 
\label{eq:secdertest}
\frac{\dif^2}{\dif t^2}f\big(M+tQ\big)\Big|_{t=0}\ge 0
\ee
in the sense of quadratic forms.  
\end{prop} 

\begin{proof} 
First one may note (by considering the case $Q=\1$) that the  assumed condition requires  $f(x)$, as a function of a real variable,  to be twice  differentiable. Taking that as granted,  
the passage from the local condition  \eqref{eq:secdertest} to \eqref{eq:conv_def} (the 
defining condition of matrix convexity)  can be deduced by noting that  
 for any given $A_0, A_1$ and $\lambda \in (0,1)$,  and any twice differentiable function $f$
\be \label{int_by_parts}
[(1-\lambda) f(A_0)+\lambda f(A_1)]-f(A_\lambda)=\int_0^1 K_\lambda(t) \frac{\mathrm{d}^2}{\mathrm{d}t^2}f(A_t)\, \mathrm{d}t,
\ee
with $A_\lambda= (1-\lambda)  A_0+\lambda  A_1\equiv A_0+\lambda(A_1-A_0)$ and 
\be 
K_\lambda(t) = 
\begin{cases}  
(1- \lambda)  t  & 0\leq t\leq \lambda \\  
(1-t) \lambda  & \lambda \leq  t \leq 1 \\ 
0 &  t \in \R \setminus [0,1] 
\end{cases}.
\ee

The relation  
\eqref{int_by_parts}  is implied through integration by parts and the observation that being continuous with 
piecewise constant derivative the function $K_\lambda(t)$ satisfies
\be \frac{\mathrm{d}^2}{\mathrm{d}t^2} K_\lambda(t) \, =\, (1-\lambda) \delta(t)  -\,  \delta(t-\lambda) + 
\lambda  \delta(t-1) 
\ee 
in the distributional sense, and vanishes beyond $(0,1)$.  (To simplify the integration by parts it is convenient to extend the integral to $[-a, 1+a]$ at an arbitrary  $a>0$.)    
\end{proof} 

Likewise, to conclude the joint convexity  
of $f(A_1,....A_k) $  
it suffices to establish the analogous version of \eqref{eq:secdertest} for function $f((M_1,....M_k) \ +\ t\, (Q_1,...,Q_k))$.  In this case the condition should be verified for arbitrary $k$-tuples $(M_1,....M_k)$ within the relevant convex domain and for arbitrary $k$-tuples of bounded self adjoint matrices $Q_j$. \\ 

As an alert to a subtlety  let us note  that while $f(\cdot)$ is a function of a single variable, when that variable is a matrix the domain's tangent space at specified $M$  is  $n^2$-dimensional.  
In  \eqref{eq:secdertest}, the multivariate nature of the local test shows up  in the requirement that the second derivative's  positivity  (in the quadratic form sense)  holds for all directions  of the linear explorations (expressed by $Q$).

 An alternative approach to the local positivity condition is to express it through the multivariate second derivative of the mapping $M\to f(M)$, as is explained and applied  in \cite{HanTom09} (in terms of 
 a generalized Hessian and  Fr\'echet differentials).   
In comparison, the condition presented in \eqref{eq:secdertest}  is simpler to state.  It also facilitates the analysis when the computation of the second derivative in  $t$ yields a recognizably positive algebraic expression.  
Examples of that are found among the functions which were successfully analyzed in  \cite{Eps73, Hia01, Lie73}, as well as those discussed here.\\

 \section{Computing with resolvents} 
\label{sec:res}

The differentiation of functions of the form $f(A(t))$ with $A(t)$ varying smoothly over $\K_n(a,b)$
 is complicated by the non-commutativity of the matrix product.  More explicitly: even if $f$ is a smooth function over $\mathbb{R}$, its value over matrices in the general case is defined through the spectral representation of the time-dependent $A(t)$, which for different times is diagonalized at different frames. Nevertheless, the test is  manageable for  a number of relevant classes of functions.  As was noted in  \cite{Eps73}, in addition to the trivial case of quadratic polynomials, simple cases include $f(x) = 1/(x-u)$ with $u\in \R\setminus (a,b)$, which  produces resolvent operators.  Through convex combinations, such   functions are the building blocks of Herglotz / Pick functions.  


\begin{lem}  \label{lem_resolvent} For any $I= (a,b) \subset R $  and $u \in \R \setminus I$, 
 the following function  is in $ \mathcal C_\infty(a,b)$
 \be 
f_u(z) =  
 \begin{cases} 
\frac{1}{z-u} & u<a \\[2ex] 
\frac{1}{u-z} & u> b  
\end{cases} 
\, \, =: \frac{\rm{sgn}_{(a,b)}(u)}{u-z} .
\ee 
\end{lem}  
\begin{proof}

Applied to matrices,  $f_u(A) = \pm (u\1-A)^{-1} $  corresponds to the  resolvent operator.  In this case the differential calculus is easy to manage through the resolvent expansion. 
To streamline its presentation, we denote   (locally within this proof)
\be
R (t):=  \frac{1}{ u\1-A(t)},
\ee
with $A(t):=A+tQ$.   

Differentiation of the resolvent to any order is facilitated by the resolvent identity: 
\be
\label{eq:resid}
\frac{1}{A+\Delta A} - \frac{1}{A}\  =\ - \frac{1}{A} \ \Delta A \  \frac{1}{A+\Delta A}.
\ee
This yields the exact and simple expression  
\be
\label{eq:secderres}
\frac{\dif^2}{\dif t^2}\ R (t)=2 R(t)\ Q\ R(t)\ QR(t)=X^*(t)\ X(t),
\ee
with $X(t): = \sqrt {R(t) } \ Q R(t)$.  The claim then follows by the clear positivity of \eqref{eq:secderres} together with Proposition~\ref{martingale_criterion}.  \\[2ex]
\end{proof}

\begin{thm}
\label{thm_2} 
For any $-\infty \leq a<b\leq +\infty$, if a function $f: (a,b)\to \R$  admits a representation of the form 
\be  \label{eq:int_rep}
f(z) = \alpha + \beta z + \gamma z^2+  \int_{-\infty}^a  \frac{(z-c)(1+uz)}{u-z}   \mu(\mathrm{d}u)  + \int_b^{\infty}  \frac{(c-z)(1+uz)}{z-u}    \mu(\mathrm{d}u) \,, 
\ee 
with $\alpha, \beta\in \!\R$, $\gamma\ge 0$, $c\in (a,b)$, and $\mu$ a finite positive measure on $\R \setminus (a,b)$, then $f$ is in $\mathcal C_\infty (a,b)$.
\end{thm} 

\begin{proof}
A function of the form \eqref{eq:int_rep}, when restricted to a compact subset of  $(a,b)$, is an integral over a bounded measure of uniformly bounded terms, corresponding to the following elementary functions (of $z$) 
\be 
\begin{split}
\label{eq:smartdiff}
\frac{(z-c)(1+uz)}{u-z}   \ &= \  \frac{(u-c)(1+u^2)}{u-z} - uz+uc-(1+u^2)  \\
&= \ (1+u^2) (u-c)\rm{sgn}_{(a,b)}(u)  f_u(z) +-uz+uc-(1+u^2).
\end{split}
\ee 
By Lemma~\ref{lem_resolvent} $f_u(A(t))$ has positive second derivative, and the same holds for their affine combination. Thus  $f\in \mathcal C_\infty(a,b)$.  
\end{proof}

The condition~\eqref{eq:int_rep}  assumed here is not unnatural since by the theorem of Kraus-Bendat-Sherman-Uchiyama~\cite{BenShe55, Kra36, Uch10} any function $f$ in $ \mathcal C_\infty(a,b)$ admits such a representation (cf.~\cite{Sim19}).   \
Thus, while Theorem~\ref{thm_2}  is one of the easiest statements to prove in the existent theory of matrix convexity, ipso-facto, it covers all $\mathcal C_\infty(a,b)$.   \\

It should be added that the collection of functions for which the  local test is to some extent manageable includes also the exponential function $f(x) = e^{\alpha x}$, though in this case the resulting algebra is a bit less elementary when the real variable $x$ is replaced by a matrix $M$.  
The exponential case played an important role in in Lieb's original concavity work \cite[Corollary 6.1]{Lie73}.  
An alternative proof based on the Herglotz / Pick representation was presented in the subsequent work of  Epstein~\cite{Eps73} and its extension by Hiai~\cite{Hia01}.

Proofs of matrix convexity through Herglotz condition, or resolvent analysis, go back to early works on the subject, including those quoted above.   Yet some novelty may potentially be found in our use of resolvent analysis for simple proofs of \emph{joint concavity} in more than one matrix variable, which is presented next.  


\section{Resolvent based proofs of joint concavity in multiple matrix variables} 


\subsection{Joint concavity of the parallel sums}  \mbox{} \\[-2ex] 
 
Turning to specific statements of interest, we next present a simple local proof   
of the concavity of the {\it parallel sums} (so called since the sum coincides with the addition rule for  resistors connected in parallel)\footnote{The parallel sum equals half of their harmonic average, and as such is one of a number of  interesting examples of operator means \cite{PusWor75}.}.   This statement was formulated and  proved  for $k=2$ by N.W. Anderson and  R.J. Duffin~\cite{AndDuf69} using an algebraic argument. 
The proof of its more general version, using the second derivative criterion, is presented here as a demonstration of the resolvent-based approach.

\begin{thm}(\cite{AndDuf69})
\label{pro:mainpro}
The mappping 
of $k$-tuples of strictly positive matrices $(A_1,\dots, A_k )$  
\be
\label{eq:newconcave}
(A_1, \dots, A_k )  \mapsto \frac{1}{A_1^{-1}+\dots +A_k^{-1}}
\ee
is jointly concave.
\end{thm}

\begin{proof}
The proof proceeds by  verifying the second-derivative criterion for the parallel sum in \eqref{eq:newconcave} with $A_j(t)= A_j+t Q_j$.  To shorten the relevant expressions, we denote (for use within this proof)  
\be
R (t):= A_1(t)^{-1}+\dots +A_k(t)^{-1} \, , 
\ee
and let $R'(t):=\dif R(t)/\dif t$, $R''(t):=\dif^2 R(t)/\dif t^2$. The existent $t$--dependence will be omitted in lengthier displayed equations. 

From the resolvent identity \eqref{eq:resid} one gets
\be
R'=-\sum_j \frac{1}{A_j}Q_j\frac{1}{A_j}, \qquad\quad R''= 2  \sum_j \frac{1}{A_j}Q_j\frac{1}{A_j}Q_j\frac{1}{A_j}\,.
\ee
We thus obtain
\begin{align}  \label{2_diff}
\frac{\dif^2}{\dif t^2}\frac{1}{R}\  &=\  2\  \frac{1}{R} \ R' \  \frac{1}{R}\  R'\  
\frac{1}{R}\  -\  \frac{1}{R}\  R''\  \frac{1}{R}  \notag \\   \notag 
&=-2\ \left( \sum_{j} \frac {1}{R} \, \frac{1}{A_j} \, Q_j \, \frac{1}{A_j} \, Q_j \, \frac{1}{A_j} \,   \frac {1}{R}   -  \sum_{j,m} \frac {1}{R} \, \frac{1}{A_j} \, Q_j \, \frac{1}{A_j}  \,   \frac {1}{R}  \, \frac{1}{A_m} \, Q_m \, \frac{1}{A_m} \,   \frac {1}{R}\right) \\[1ex] 
&=-2 \, \sum_{j,m} Y_j^* \left( \delta_{j,m} -T_{jm}\right) Y_m ,
\end{align}
where
\be
Y_j:=\frac{1}{\sqrt{A_j}}Q_j\frac{1}{A_j}\frac{1}{R}, \qquad\quad T_{j,m}:=\frac{1}{\sqrt{A_j}} \frac{1}{R}\frac{1}{\sqrt{A_m}}\,.
\ee  
Viewing  $T=[T_{j,m}]$ as a matrix of operators (matrices),  we note that it has the self-adjoint projection properties: 
\be 
 T^{\dagger} = T\,, \quad T^2 = T  \quad (\mbox{in the sense that $\sum_k T_{j,k}\ T_{k,m}  = T_{j,m}  $})   \,.
\ee
Hence $\1- T =  [\1 - T]^{\dagger} \, [\1-T]$.  
 It follows that the expression in the last line of \eqref{2_diff} is a negative matrix, and thus the concavity criterion of  Proposition~\ref{martingale_criterion} is met. 
 \end{proof}

\subsection{Lieb concavity} \mbox{} \\[-2ex] 

From the concavity of parallel sums one may deduce the following result on the joint concavity of powers'  tensor  products.  This statement forms a particular case of Corollary 6.1 in Lieb's \cite{Lie73}.  
In \cite{Hia01} one finds such a result proven using Pick function analysis (Corollary 2.2 there), and also a more complete discussion of statements in this vein.

\begin{thm} 
\label{thm_5}
For any  real numbers  $p_1, ..., p_k$ obeying 
\be 
p_j \geq 0,\, \qquad \sum_{j=1}^k p_j  \leq 1 \,. 
\ee 
the mapping 
\be \label{eq_prod1} 
(A_1, A_2, ..., A_k) \mapsto A_1\,^{p_1} \otimes  A_2\,^{p_2} \otimes  ... \otimes A_k\,^{p_k}
\ee  is jointly concave on $k$-tuples of  strictly positive matrices.
\end{thm}

\begin{proof} 

 It is  convenient to present the tensor product  as a regular product of commuting operators,  
\be \label{eq_prod}
A_1\,^{p_1}\otimes  A_2\,^{p_2} \otimes  ... \otimes A_k\,^{p_k} = \
  \widetilde A_1\,^{p_1} \cdot   \widetilde A_2\,^{p_2}\cdot ... \cdot  \widetilde  A_k\,^{p_k}
\ee 
with $ \widetilde A_j := \1 \otimes \dots  \otimes \1 \otimes A_j  \otimes \1 \otimes  \dots \otimes \1$.

Let us first prove the claim for  sequences of strictly positive numbers  with $\sum_{j=1}^k p_j =1$,  $k\geq 2$.   In that case, for any collection of commuting matrices $\widetilde A_j $
\begin{multline} \label{A_integral_rep}
 A_1\,^{p_1}\otimes  A_2\,^{p_2} \otimes  ... \otimes A_k\,^{p_k} = \
  \widetilde A_1\,^{p_1} \cdot ... \cdot  \widetilde  A_k\,^{p_k} \\ 
  =  C_k(p_1, p_2,\dots,p_k)^{-1} \int_0^\infty \cdots \int_0^\infty 
 \left(
  \frac{1}{\widetilde A_1} + \frac{u_2 }{\widetilde  A_2} + \dots + 
\frac{u_k }{\widetilde  A_k }  \right)^{-1} 
  \, \prod_{j=2}^k \left(u_j^{p_j}\,\frac{\mathrm{d}u_j}{u_j}\right) \,. 
  \end{multline}
  with the finite constant
  \be 
 \label{eq:usfconst}
C_k(p_1,\dots,p_k):=\int_0^\infty \cdots \int_0^\infty \frac{1}{1+u_2+\dots+u_m}\, \prod_{j=2}^k \left(u_j^{p_j}\,\frac{\mathrm{d}u_j}{u_j}\right) \,.  
\ee

The  convergence of the above integral 
 (which is suggested by simple power counting)  can be seen  by  transforming it  and changing the variables of integration to $v_j:=t \,u_j$  as follows
 \begin{multline}
 C_k(p_1,\dots,p_k) = \int_{\R_+^{k}}   e^{-t(1+u_2+\dots+u_k)}\,  \prod_{j=2}^k \left(u_j^{p_j}\,\frac{\mathrm{d}u_j}{u_j}\right) \mathrm{d}t\,\ =\  
 \prod_{j=1}^k 
 \left(\int_0^\infty e^{-v_j}  v_j^{p_j}\,\frac{\mathrm{d}v_j}{v_j} 
  \right) \  < \ \infty  \,.
\end{multline}

By Theorem~\ref{pro:mainpro} for each $\{u_1,..., u_k\}$ the integrand in \eqref{A_integral_rep} is 
a jointly concave function of $\{A_1,..., A_k\}$.   Combined with the uniform convergence of the integral  over  compact sets of  $k$-tuples of $A_j$   
this allows to deduce the concavity of the tensor product which is on the left side of \eqref{A_integral_rep}.

The statement's extension to $\sum_{j=1}^k p_j <1$ can be deduced from the case $k+1$ with $\sum_{j=1}^{k+1} p_j =1$, upon setting $\widetilde A_{k+1} \equiv \1$. 
Once that is established, the case where some $p_j=0$ easily follows from the validity of this statement  for $\sum_{j=1}^k p_j <1 $ with strictly positive $j$, at  smaller values of $k$.  
\end{proof} 

In addition to its intrinsic value  Theorem~\ref{thm_5} is of interest  as a gateway (or one of such)  to the following theorem of E.H. Lieb, for which an elementary proof is presented below.  

\begin{cor} (~\cite{Lie73}) \label{cor_AB} 
For each fixed matrix $K$,  and $p, r \in [0,1]$ with $p+r \leq 1$, the following function of two matrix variables 
\be\label{Lieb_concavity}
(A,B)  \mapsto \mathrm{Tr} \big[A^p K^* B^r K\big]
\ee 
is jointly concave over $\K_n(0,\infty)\times\K_n(0,\infty)$.     
\end{cor} 
\begin{proof} 
The latter follows from the $k=2$ case of Theorem~\ref{thm_5} through the observation that for any $K : \mathcal H_1 \mapsto \mathcal H_2  $  there exists a vector $\ket{ \mathcal K} \in \mathcal H_1 \otimes  \mathcal H_2 $ (in Dirac's notation) such that 
for all $A:   \mathcal H_1 \mapsto \mathcal H_1  $ and $B:   \mathcal H_2 \mapsto \mathcal H_2  $
\be
\mathrm{Tr}[A^pK^*(B^t)^rK\big]= \bra{ \mathcal K}  A^p\otimes B^r \ket{ \mathcal K}   \, . 
\ee
\end{proof} 


An immediate implication of Lieb-concavity, Corollary~\ref{cor_AB}, and the context in which the result appeared first, was Lieb's proof of the Wigner-Yanase-Dyson conjecture \cite{Lie73}.    
It states that the Wigner-Yanase\cite{WigYan63} ``p-skew information'' of a density matrix $\rho$, with respect to an operator $K$,  is  concave in $\rho$.    
This quantity is defined as 
\be 
I_p(\rho, K) := \frac 12 \mathrm{Tr} [\rho^p, K^*] [\rho^{1-p}, K] \,\,  \left(= \mathrm{Tr} K \rho^p K \rho^{1-p}  - Tr K \rho K \right) ,
\ee 
where $[A,B] := AB-BA$ denotes the commutator.  Its convexity as a function of $\rho$ follows readily from Corollary~\ref{cor_AB} with $A=B=\rho$ and $r=1-p$.

\subsection{The Kubo-Ando Theorem} \mbox{ } \\[-2ex]

Theorem~\ref{thm_5} yields also a streamlined proof of the following theorem of T. Kubo and T. Ando \cite{KubAnd80}.   A simplification of the original derivation was presented by E.G. Effros~\cite{Eff09} using the Hansen-Pedersen Jensen operator inequality (\cite{HanPed03}).   The proof given below takes a short-cut through Loewner's classification theorem, but given that it may seem shorter.


\begin{thm}[\cite{KubAnd80}]
\label{thm_3}
If  $f:(0,\infty)\to \mathbb{R}$  is operator monotone, then for any $n$ the function
 \begin{equation}
 \label{eq:prosp}
 (A,B)\mapsto B^{1/2}f\big(B^{-1/2}AB^{-1/2}\big)B^{1/2}
 \end{equation}
 is operator convex on $\K_n(0,\infty)\times \K_n(0,\infty)$.
\end{thm}


\begin{proof}

By Loewner's representation theorem of matrix monotone functions ~\cite{Loe34} the above function of $(A,B)$ can be presented as 
\be
B^{1/2}f\big(B^{-1/2}AB^{-1/2}\big)B^{1/2}=aA+bB+\int_0^\infty \frac{1}{(tA)^{-1}+B^{-1}} \cdot\frac{1+t}{t} \mathrm{d}\nu(t)
\ee
at some $a,b\in \R$, and $\nu$ an appropriate measure on $(0,\infty)$ (cf. \cite[Theorem 3.1]{Sim19}). The stated concavity therefore  follows from the concavity of parallel sums, of Theorem~\ref{thm_5}.
\end{proof}

 \section{Applications to quantum entropy} 
 \label{sec:entropy}
 
\subsection{Basics  notions} \mbox{} \\[-2ex] 

 Convexity considerations are of fundamental significance in statistical mechanics.   In particular that is so in relation to the entropy function -- a term which shows up in a number of different contexts, with slightly varying meaning and properties, cf. \cite{Lie75}.  \\

In quantum physics the states of a system are presented as expectation value functionals defined over the  self adjoint (bounded) operators in the corresponding separable Hilbert space $\mathcal H$.    These  are  described by density operators $\rho$, which are positive and of trace $\mathrm{Tr} (\rho) =1$, in terms of which the expectation value functional is the mapping 
 \be  A \mapsto \mathrm{Tr} (A \rho).
\ee 
The von Neumann entropy of the state is defined as 
\be 
\label{eq:entropy}
S(\rho) := - \mathrm{Tr} (\rho \log \rho)   \equiv \mathrm{Tr} F(\rho)
\ee 
 where $F$ is  the function  $F(x) = - x \log (x) $.  \\ 
 
  The basic notions and fundamental properties of entropy are explained in a pedagogical manner in   \cite{Lie75, LieRus73}. The entropy defined in \eqref{eq:entropy} is non-negative, vanishes if and only if $\rho$ is a rank-one projection, and it attains its maximal value at $\rho = \frac{1}{\dim \mathcal H} \1$. \\ 

For systems composed of two disjoint, but possibly interacting, components the relevant Hilbert space is a tensor product, 
$\mathcal H_{12} = \mathcal H_1\otimes \mathcal H_2 $.
A general state on such a system would be referred to as $\rho_{12}$. Its restriction to a sub component, e.g. observables of the form 
$A \otimes \1$ defines the state operator $\rho_1$\,.

The mapping 
 \be 
 \rho_{12} \mapsto \rho_1
 \ee
can be accomplished through partial trace, or alternatively through the following relation
\be \label{U_trace}
\int (\1 \otimes U ^*) \, \, \rho_{12} \, \, ( \1 \otimes U) \,\, \nu_2(\mathrm{d}U)  =  \rho_{1}\otimes \frac{1}{\mathrm{dim}   \mathcal H_2 }  \1
\ee 
where $\nu_2$ is the normalized Haar measure on the group of unitary transformations on $\mathcal H_2$. \\

Another useful mapping is the de-correlation of the two components (or ``pinching'' - in the terminology of Ch. Davis \cite{Dav59, SutBerTom17}).  In presenting and applying it here we correct the mistake in the paper's published version's Eq (5.7). \\ 

\noindent{\bf Note:}   To avoid confusion between versions, the modified equations are marked here by an asterisk ($'$), and the bulk of the modified text is bracketed by $<<<$ ... $>>>$.  \\ 

%

$<<<$
Given a state operator $\rho_{12}$ on $\mathcal H_{12} = \mathcal H_{1} \otimes \mathcal H_{2}$, let $\widehat \rho_{12}$ be the operator which is diagonal in the tensor product's orthonormal basis consisting  of  functions  $ |k,l\rangle_{12}   = |k\rangle_1 \otimes |l\rangle_2 $,  where  $|k\rangle_1$ and $|l\rangle_2$ are normalized eigenvectors of $\rho_1$ and $\rho_2$, 
with the matrix elements 
\be \tag{5.5'}
 \langle k,l| \widehat{\rho}_{12}|l',k'\rangle= \delta_{kk'}\delta_{ll'}(\rho_{12})_{kl,k'l'} \, . 
 \ee
   
The mapping $\rho_{12} \mapsto \widehat \rho_{12} $ can be accomplished through a probability average over transformations of $\rho_{12}$ under unitary mappings.  
For that, let $\nu_{\rho_{12}}(\mathrm{d}U)$ be the Haar probability measure over the group of unitaries which in the above basis acts as multiplication operators  of independently varying phases 
\be \tag{5.6'}
\mathcal {G}_{\rho_{12}} =  \{ \prod_{\substack {k =1,..., \dim(\mathcal H_1) \\   l =1,..., 
\dim(\mathcal H_2) }} e^{i\theta_{k,l}}\} \,. 
\ee
In these terms, $\widehat{\rho}_{12}$ is presentable as the ``pinched state'' : 
\begin{equation} \tag{5.7'} 
\label{eq:doublepinching}
\widehat{\rho}_{12} =\int U^*\rho_{12} U\, \nu_{\rho_{12}}(\mathrm{d}U).
\end{equation}

The mapping $\rho_{12} \mapsto \widehat{\rho}_{12}$ 
 is easily seen to have the following properties: 
\begin{enumerate}[1)]
\item 
The states $\widehat{\rho}_{12}$  and $\rho_{12}$ coincide on the commuting algebra of 
functions $F(X_1,X_2)$ of the  operators $X_1=\sum_k k \cdot \ket{k}\bra{k}\otimes \1$ and $X_2=\sum_l l \cdot \1\otimes \ket{l}\bra{l}$.   
\item Being diagonal in the above basis, the state $\widehat{\rho}_{12}$ is presentable as a classical probability distribution of a pair of random variables of values corresponding to $l$ and $k$, with $\Pr\left(\{(X_1,X_2) = (k,l)\}\right)= (\rho_{12})_{kl,kl}$.  
\item The restrictions of the two states on each of the two subsystems coincide, that is:
\be \label{5.8'}  \tag{5.8'}
\widehat \rho_1 = \rho_1\, , \quad \mbox{and}\quad \widehat \rho_2 = \rho_2\, .  
\ee  
\end{enumerate} 
 $>>>$\hfill  \\	

\setcounter{equation}{8} 

 In discussing the entropies of the states induced on subsystems, it is customary to abbreviate
 \[
 S_1=S(\rho_1), \qquad  S_{12}=S(\rho_{12}), \qquad  S_{123}=S(\rho_{123})\,.
\]
 
 In the analogous discrete classical  systems, in each state $\rho$  the entropies of the states induced on subsystems are increasing in the subsystem size, e.g. 
\be    S_1 \leq S_{12}.
 \ee 
This feature clearly does not extend to the quantum case, e.g. a composite system can be in a pure state of zero entropy while its subsystems will not be so.   

Although the analogy with the classical case in places breaks down, it is  of value as a guide. Among the first generally valid properties of the above constructs is the \emph{subadditivity of entropy}:

 \begin{thm} (Subadditivity) For any state of a composite  system
 \be  \label{SE}
 S_{12}  \leq S_1+S_2.
 \ee 
 \end{thm}
\begin{proof}
The statement can be reached through the following chain of relations 
\be \label{subadd} \tag{5.11'}
S( \rho_{1,2}) \  \leq  \  S\left(\widehat{\rho}_{12} \right) \le S(\widehat{\rho}_1)+S(\widehat{\rho}_2)=S(\rho_1)+S(\rho_2) \,,
\ee
Here, the first step follows from \eqref{eq:doublepinching} by the Jensen inequality,
 the second inequality is by the classical subadditivity of entropy, and the equality holds by \eqref{5.8'}.
\end{proof}
\setcounter{equation}{11} 

Entropy's  subadditivity  \eqref{SE} can be read as saying that 
the increment 
\be  S(1\vert 2) :=  S_{12} - S_{2} \,, 
\ee 
to which we refer as the conditional entropy of component $1$ given $2$,  satisfies
\be 
S(1\vert 2)  \leq  S_{1} \,.  
\ee

In the analogous classical discrete systems the state of a composite system is described by a probability distribution with weights 
$\rho(\omega_1, \omega_2,...)$ over the space of configurations $\Omega=\Omega_1\times \Omega_2\times ...$. 
The probability distribution of $\omega_1$, may be presented as the average of  the conditional distribution $\rho(\omega_1\vert \omega_2)$.   A simple calculation then shows that for classical systems  $S(1\vert 2)$ equals  the average over  $\omega_2$ (with measure $\rho_2$) of the entropy of the conditional probability distribution $\rho(\omega_1\vert \omega_2)$.  
From this emerged the term   {\it relative entropy}.   \\ 

\noindent{\bf Remark} 
We also point out that  the above considerations lead to  a decomposition of potential interest of the  quantum mutual information between two subsystems,   defined as 
 \[ \tag{I.1}
I_\rho(1:2) : = S(\rho_1) + S(\rho_2) - S(\rho_{12})  \, .
\] 
The inequalities presented in \eqref{subadd} allow us to split the above difference into a sum of two positive terms, of which the first   is of quantum nature and the second is purely classical: 
\[  \tag{I.2}
I_\rho(1:2) = \left[ S(\widehat{\rho}_{12})  - S(\rho_{12}) \right] + \left[  S(\rho_1) + S(\rho_2) -  S(\widehat{\rho}_{12}) \right] 
\] 
The first summand is the increase in the quantum state's entropy which results from the measurement of the pair of commuting observables $(X_1, X_2)$ (adapted to  $\rho$).
The second summand  is the increase in the classical entropy when a joint distribution of a pair of classical variables is replaced by the (uncorrelated) product of its marginal probability distributions.  

We should however hasten to add that  decompositions into classical versus quantum parts of correlations are not unique.  In particular, the one explained above differs from the two that were introduced in \cite{Hen_Ver01, Oli_Zur01},  guided by considerations of  information transmission and state teleportation.    \\

If there is a third component, from the above description through a judicious application of the Jensen inequality one may conclude that in the classical case 
\be 
S_{123}-S_{23} \leq S_{12}-S_2  \,. 
\ee  
This relation is referred to as the \emph{strong subadditivity of entropy (SSA)}.
Its intuitive interpretation is that as each measurement adds information  (in the classical case), the incremental entropy is decreasing.    \\ 


The above reasoning does not apply to quantum systems.  Nevertheless, O. Lanford and D. Robinson~\cite{LanRob68} conjectured in 1968 that  SSA is valid also in the quantum case.  
The conjecture was proven in 1973 in the joint work by  E.H. Lieb and M.B. Ruskai~\cite{LieRus73}.  
Already in that paper the authors provide a number of proofs.  Given the interest in the subject, it is not surprising that since then a number of simplifications have emerged.   Following is a streamlined proof using the results presented above.  

\subsection{The Lieb-Ruskai concavity} \mbox{ } \\[-2ex]

A key step  in the first listed proof of SSA in \cite{LieRus73}  is the following noteworthy statement (Theorem 1 in \cite{LieRus73}), which is also of independent interest, cf. \cite{Han16}. 

\begin{thm}[Lieb-Ruskai \cite{LieRus73}] \label{LR_concavity}
In any composite system the conditional-entropy 
\be \label{rho_to_cond_entropy}
\rho_{12} \mapsto S(\rho_{12}) - S(\rho_{2}) \equiv  \mathrm {Tr}  \rho_{12} \big[ \log \rho_{12} -   \1 \otimes \log  \rho_2 \big].
\ee 
is a concave function of the state operator $\rho$.  
\end{thm} 

The proof (version of G. Lindblad~\cite{Lin74}) starts with a slightly more general version of the statement,  asserting that the following mapping is jointly concave over pairs of positive matrices
\be\label{logA_logB}
(A,B) \mapsto \,   - \,\, \mathrm{Tr} \big[ A  (\log A -\log B ) \big] \,. 
\ee
The combination on the right is referred to as the relative entropy, and denoted 
\[
S(A\vert B) := - \mathrm{Tr} \big[ A  (\log A -\log B ) \big]
\]  
\begin{proof}
By Lieb concavity, stated above as Corollary~\ref{cor_AB}, for every $\varepsilon >0$ the following mapping is jointly concave over pairs of positive matrices
\be
(A,B) \mapsto \frac{\mathrm{Tr} _{12}\big[ A^{1-\varepsilon}  B^\varepsilon - A\big]}{\varepsilon}  \equiv 
\mathrm{Tr} _{12}\big[ \frac{A^{1-\varepsilon} -A}{\varepsilon} B^\varepsilon +  A \frac{ B^\varepsilon - \1}{\varepsilon} \big].
\ee 
The claim follows by taking the limit $\varepsilon \searrow 0$.

To relate this with \ref{rho_to_cond_entropy},  apply the above to   
 $A= \rho_{12}$, $B = \1\otimes \rho_2$.  In this  case   
\begin{eqnarray} 
\mathrm{Tr} \big[ A  \log B ) \big] &=&  
 \mathrm {Tr}_{12}\,\, \rho_{12} \,\, \log  (\1 \otimes \rho_2)= \mathrm {Tr}_{12}\,\, \rho_{12} \,\,  (\1 \otimes  \log \rho_2) \notag \\[1ex]  &=& \mathrm {Tr}_{2} \,\,\rho_{2}  \log  \rho_2 = -S_2\,, 
\end{eqnarray}
where the subscript on $\mathrm {Tr}$ indicates the Hilbert space over which the trace is performed.  Thus, under this substitution  
\be
S(A\vert B)=  - \mathrm {Tr}  \rho_{12} \big[ \log \rho_{12} -   \1 \otimes \log  \rho_2 \big]
 = 
S(\rho_{12}) - S(\rho_{2}) \,. 
\ee 
Since both $A$ and $B$ are linear functions of $\rho$ the just proven  jointly concavity of $S(A\vert B)$    implies the claimed concavity of the conditional entropy as function of  $\rho$. 
\end{proof} 

\subsection{The strong subadditivity of quantum entropy} \mbox{} \\[-2ex]  

\begin{thm}[\cite{LieRus73}]  The von Neumann entropy of a quantum (finite dimensional) systems is strongly subadditive, in the sense that for any state of a composite system
 \be \label{eq_SSA}
 S(\rho_{123}) - S(\rho_{23})  \leq  S(\rho_{12})-S(\rho_2)\, .
 \ee 
\end{thm}
Following is a streamlined deduction of this statement from the Lieb-Ruskai concavity, incorporating an argument of A. Uhlmann~ \cite{Uhl73}[Section 8]. 
\begin{proof}   

As seen in \eqref{U_trace},
any state $\rho_{123}$ can be transformed into one in which the third component is independent from the first two through a probability average over unitary transformations which do not affect the distribution of the first two components ($U=\1\otimes \1 \otimes U_3$).   
The resulting  state 
\be \tag{5.21'}
\widetilde \rho_{123} =  \rho_{12}\otimes  (\mathrm {dim} \mathcal H_3)^{-1} \1   
\ee
is the probability average of states whose  relative entropy $S(1\vert 2)$ equals that of $\rho_{123}$.   
Thus from the Lieb-Ruskai concavity principle (Theorem~\ref{LR_concavity}), combined with the Jensen inequality \eqref{Jensen}, it follows that 
\setcounter{equation}{21} 
\be 
S( \rho_{123}) - S( \rho_{23}) \leq S(\widetilde \rho_{123}) - S(\widetilde \rho_{23}) \,.
\ee 
However, due to the above product structure of  $\widetilde \rho_{123}$  
the contribution of the third component to its entropy is strictly additive, while its restriction to the first two components equals that of $\rho$.  Therefore 
\be 
S(\widetilde \rho_{123}) - S(\widetilde \rho_{23})  = S(\widetilde \rho_{12}) - S(\widetilde \rho_{2})  = 
 S(\rho_{12}) - S(\rho_{2})\,.  
\ee 

\end{proof}

\appendix  
 
Let us add that while our discussion is carried in the context of matrices of finite order $n$,  by continuity arguments,  spelled in \cite{BenShe55} and \cite{LieRus73}(appendix by B. Simon),   results that hold uniformly in $n$ admit natural extensions to 
bounded linear operators over separable Hilbert spaces, i.e. the class $\K_\infty(a,b)$.

It should also be noted that the above-cited works on quantum entropy,   and other contributions by mathematical physicists, spawned a growing body of interesting results concerning matrix convexity.  
A relevant recent review can be found in \cite{Car22}.  
 
 \bigskip

	\noindent\textbf{Acknowledgements:}
We thank Ramon van Handel and Simone Warzel for useful comments on an earlier draft of this manuscript and Benjamin Bobell for motivating discussions.  M.A.  gratefully acknowledges the support of the Weston Visiting Professorship at the Weizmann Institute of Science (Rehovot, Israel), on a visit during which some of the work was done.	\\




\end{document}